\documentclass[aip,jcp]{revtex4}
\usepackage{comment}

\usepackage{amsmath,amssymb,graphicx}
\usepackage{epsfig}
\usepackage{graphicx}
\usepackage{enumerate}

\newcommand{\beq}{\begin{equation}}
\newcommand{\eeq}{\end{equation}}

\newcommand{\bea}{\begin{eqnarray}}
\newcommand{\eea}{\end{eqnarray}}

\begin{document}
\title{
Two-level system in spin baths: Non-adiabatic dynamics and heat transport
}
\author{Dvira Segal}
\affiliation{Chemical Physics Theory Group, Department of Chemistry, University of Toronto,
80 Saint George St. Toronto, Ontario, Canada M5S 3H6}

\date{\today}
\begin{abstract}
We study the non-adiabatic dynamics of a two-state subsystem in a bath of independent spins using the non-interacting blip
approximation, and derive an exact analytic expression for the relevant memory kernel. We show that
in the thermodynamic limit, when the subsystem-bath coupling is diluted (uniformly) over many (infinite) degrees of freedom,
our expression reduces to known results, corresponding to the harmonic bath
with an effective, temperature-dependent, spectral density function.
We then proceed and study
the heat current characteristics in the out-of-equilibrium spin-spin-bath model,
with a two-state subsystem bridging two thermal spin-baths of different temperatures.
We compare the behavior of this model to the case of a spin connecting boson baths,
and demonstrate pronounced qualitative differences between the two models.
Specifically, we focus on  the development
of the thermal diode effect, 
and show that the spin-spin-bath model cannot support it at weak (subsystem-bath) coupling,
while in the intermediate-strong coupling regime its rectifying performance outplays the spin-boson model.

\end{abstract}

\maketitle

\section{Introduction}
\label{Intro}

Complex aspects of quantum dynamics in condensed phases may be captured within simple models of few-state
subsystems immersed in dissipative environments \cite{Weiss, Mahan, Nitzan,Legget}. In the
``Caldeira-Leggett" model \cite{Caldeira-Legget} the thermal bath includes a collection of independent
harmonic oscillators with linear coupling to the subsystem. It allows for the derivation of the quantum
Langevin equation, useful e.g. for studying quantum barrier crossings in complex environments \cite{Nitzan}.
A particularly interesting Caldeira-Leggett-type problem is the spin-boson model, comprising a two-state
subsystem. This minimal model can describe the dynamics of a single charge on two molecular states coupled to a
dissipative bath (solvent) \cite{Weiss}, or the Kondo problem for magnetic impurities \cite{Kondo}. Since this model
is not solvable analytically, an array of treatments, analytical and numerical, have been developed for
evolving the two-state dynamics. A non-exhaustive list includes perturbation theory approaches in the subsystem-bath interaction or the
non-adiabatic parameter \cite{Weiss,Legget,Harris,Aslangul,Dekker}, mixed quantum classical and
semiclassical methods \cite{mixed1,Reich1}, path integral tools \cite{egger1,egger2,QUAPI}, numerical
renormalization techniques \cite{NRG1,RevRG} and numerically-exact wavefunction schemes
\cite{MCTDH1,MCTDH2}.

The spin-boson model is of interest beyond the question of the reduced (subsystem) dynamics. 
The simple picture can be extended 
to the out-of-equilibrium regime by coupling the central spin to two separate (spatially or energetically) thermal reservoirs which are maintained at different temperatures.
This setup has been proposed as a toy model for exploring quantum
transport phenomenology in an anharmonic nanojunction \cite{SegalRectif,SegalM}. Given the complex
dissipative dynamics observed in the single-bath model, it is not surprising that the related
out-of-equilibrium version exhibits a rich transport behavior \cite{Thoss1,Thoss2,Saito,SegalQ}. We have recently
extended the non-interacting blip approximation \cite{Legget,Dekker,Aslangul}
and studied heat transport characteristics in the two-bath spin-boson model in the non-adiabatic regime \cite{SegalRectif,SegalM,FR2}:
We have validated the heat exchange fluctuation theorem in this many-body model and demonstrated
that the heat current can display a significant negative differential thermal conductance (NDTC). 
To understand heat transport in actual nanostructures, it is of
interest to explore transport behavior between other reservoirs, particularly considering anharmonic
environments. While the reduced dynamics may conceal the nature of the bath \cite{Kosloff}, in this work we
show that in an out-of-equilibrium situation the steady-state heat current displays marked,
qualitative contrasts, when the subsystem is attached to distinct environments.

Harmonic baths serve for describing collective (normal) modes of the media: phonons
and photons. To include anharmonicities  \cite{Kosloff,MakriAn,ThossAn} and account for other relevant
media with localized modes, spin glasses or magnetic materials,
a different reservoir has been proposed: a spin-bath
model \cite{CaldeiraSpin,Stamp}. In its simple form this reservoir includes a collection of non-interacting spins,
and it may be viewed as the extreme anharmonic limit of the harmonic bath, with each vibration truncated
to its lowest two states.

A detailed comparison between the dissipative dynamics of a two-level
subsystem (TLS) immersed in either harmonic baths \cite{Legget} (spin-boson model) or spin baths \cite{Stamp} (central spin model or
spin-spin-bath model)
has been carried out in several works.
The problem has been addressed
using the resolvent operator approach \cite{Hanggi}, and in the non-adiabatic limit with the non-interacting blip
approximation (NIBA)  \cite{MakriSpin}. Numerically-exact simulations were performed using various tools: iterative
path integral methods \cite{MakriSpin}, the multilayer multiconfiguration time-dependent Hartree theory
(ML-MCTDH) \cite{Wang}, and the surrogate Hamiltonian approach \cite{Kosloff}. Bloch-Redfield equations,
perturbative in the subsystem-bath interaction energy, were presented in Ref. \cite{Lu}.
Specifically, it has been shown in several works \cite{Hanggi,MakriSpin,Wang} that in the non-adiabatic
regime the subsystem coherent-incoherent crossover is shifted to stronger coupling when increasing the
temperature of the attached spin bath. In other studies it has been argued that harmonic
oscillators and spin baths lead to a similar reduced dynamics (energy relaxation, decoherence,
entanglement) in the weak-to-intermediate coupling regime \cite{Kosloff}. Overall,  while
differences exist, the reduced dynamics does not readily evince on the physical nature of the environment. It
is important to find whether other dynamical or steady-state properties may more clearly attest to the
intrinsic properties of the dissipative bath.

The spin-spin-bath model can be mapped onto the spin-boson model when the
spectral density function is continuous and the subsystem-bath coupling is diluted over many bath modes
\cite{Silbey,Makriproof}. Earlier studies with NIBA assumed this to be the case \cite{Hanggi,MakriSpin,Wang}.
However, as discussed in details in Ref. \cite{Wang},
this ``linear response" mapping should be taken
with great caution since many bath modes should be included in the simulation to approach the thermodynamic limit, particularly in the
non-adiabatic limit when the spectral function is broad or when the temperature is low \cite{Makriproof,Wang}.
 Moreover, the mapping is invalidated when the bath
includes localized modes which are not weakly coupled to the central spin, even in the thermodynamic limit \cite{Stamp}.

Here we provide an exact expression for the memory kernel of the NIBA 
in the case of a spin bath. It allows one to study the
non-adiabatic dynamics of a central spin in more general and realistic (finite, strongly-coupled)
environments of localized modes.
Naturally, one could ask why bother with NIBA, an approximate theory,  when exact numerical tools as described e.g. in Refs.
\cite{MakriSpin,Wang} are available. The answer is that for an $N$-state subsystem NIBA equations
include $N^2$ elements. In contrast,  a numerically-exact approach
such as the iterative implementation of the quasi-adiabatic path integral expression \cite{QUAPI,MakriSpin}
(formidably) scales as $N^{2K}$,  $K$ is an integer satisfying
$K= \tau_M/\delta t$, where $\tau_M$ is the bath decorrelation time and $\delta t$ the simulation time step.
Thus, one should advance and improve approximate theories  hand-in-hand with the development of exact simulation tools.

Following our derivation of NIBA and its kernel, and the clarifications on the assumptions behind its linear-response-thermodynamical
limit, we proceed and study the heat transport behavior in a junction made of a TLS coupled to two thermal baths of spins.
In this case, the subsystem's
dynamics is obtained by solving a time-convolution quantum master equation, with a kernel that is
non-additive in the baths' degrees of freedom. As a result,
the heat current involves compound, non-additive, relaxation and excitation processes in both baths.

This work includes two contributions. In the first part of the paper, Sec. \ref{Model}, we derive the
NIBA equations for a spin bath, providing an exact expression for the memory kernel.
In the so-called ``thermodynamic limit" it
reduces to known results \cite{Makriproof}.
In the second part of our work, Sec. \ref{Transport},  we consider the thermodynamic limit of the NIBA and
present numerical results for the heat current through a TLS mediating either harmonic baths
or spin baths.
In what follows we set the Boltzmann constant as $k_B=1$, and work in units of $\hbar=1$.

\section{Model and Method}
\label{Model}

\subsection{Spin-spin-bath model}

We consider a single spin (subsystem) and couple it to two baths ($\nu=L,R$)
prepared at different temperatures $T_{\nu}=\beta_{\nu}^{-1}$.
Each bath includes a collection of non-interacting spins, and
the total 
Hamiltonian is written as
\bea
H_{SS}&=&\frac{1}{2}\omega_0 \sigma_z +\frac{1}{2}\Delta \sigma_x +
\frac{1}{2}\sum_{\nu,j}\omega_j \sigma_z^{\nu,j}
\nonumber\\
&+&\frac{1}{2}\sigma_z\sum_{\nu,j}\lambda_{\nu,j}\sigma_x^{\nu,j}.
\label{eq:H}
\eea
Here $\omega_0$ is the energy gap between the states of the central spin, $\Delta$ is the tunneling
element, and $\sigma_{x,y,z}$ are the Pauli matrices of the subsystem. Similarly,
$\sigma^{\nu,j}_{x,y,z}$ are the Pauli matrices of the $j$th spin in the $\nu$ bath, coupled at strength
$\lambda_{\nu,j}$ to the polarization of the central spin. For simplicity,
 $\lambda_{\nu,j}$ is assumed real, but generalizations beyond that are trivial.


The spin-boson Hamiltonian can be transformed via the small polaron transformation into the basis of
displaced oscillators \cite{Mahan}, see a short description in Appendix A. Similarly, since the
particles in the spin bath are non-interacting, Eq. (\ref{eq:H}) can be readily transformed to
the basis of displaced spins. The unitary transformation $\tilde H_{SS}=e^S (H_{SS}) e^{-S}$, $S=S_L+S_R$,
diagonalizes the last two terms in Eq. (\ref{eq:H}). The generator of the transformation is \cite{Lu}
\bea
S_{\nu} = \frac{i\sigma_z} {2} \sum_j \arctan(\eta_{\nu,j})\sigma_{y}^{\nu,j},
\label{eq:S}
\eea
with the dimensionless parameter
\bea
\eta_{\nu,j}=\frac{\lambda_{\nu,j}}{\omega_j}.
\eea
The Hamiltonian now reads
\bea
\tilde H_{SS}=
\frac{1}{2}\omega_0 \sigma_z  + \frac{1}{2}\sum_{\nu,j}\tilde \omega_{\nu,j} \sigma_z^{\nu,j}
+\frac{1}{2}\Delta \left[\sigma_+e^{i\Omega} +\sigma_-e^{-i\Omega}\right],
\label{eq:Htilde}
\eea
where $\sigma_{\pm}=\frac{1}{2}(\sigma_x\pm i \sigma_y)$ are the auxiliary Pauli matrices.
The bath operators are given by
\bea
\Omega=\sum_{\nu,j} \arctan (\eta_{\nu,j}) \sigma_{y}^{\nu,j},
\label{eq:Omega}
\eea
and the spin frequencies, now corrected by the subsystem-bath coupling parameter,  become
\bea
\tilde\omega_{\nu,j} = \omega_j \sqrt{ 1+\eta_{\nu,j}^2}.
\label{eq:omegatilde}
\eea
%
In what follows, we refer to the two-bath spin-spin-bath model (\ref{eq:H}) as the ``SS model".
Similarly, the two-bath spin-boson model (\ref{eq:HH}) is identified as the ``SB model".

\subsection{NIBA equations}

We employ the NIBA scheme which is well established for describing the non-adiabatic dynamics
of the single-bath spin-boson model: It can reproduce the correct subsystem dynamics at
strong interactions or at high temperatures when the bath is Ohmic. It is
also exact for the case of a degenerate (unbiased) subsystem, 
at weak damping \cite{Weiss}. 
More recently, NIBA has been tested for the (single-bath) spin-spin-bath model.
A detailed analysis of its accuracy against ML-MCTDH simulations, at zero temperature in the unbiased case,
has been carried out in Ref. \cite{Wang}. It was found that NIBA could qualitatively capture the
coherent-incoherent transition at zero temperature, and that it became exact in the scaling limit. At
higher temperatures, NIBA is expected to have a better performance, to provide a quantitative agreement with exact
simulations.

We have recently extended the NIBA  to the {\it out-of-equilibrium} two-bath case, considering harmonic
environments of different temperatures \cite{SegalRectif,SegalM,FR2}. Next we further point out that this
extension holds for spin environments.
More fundamentally, considering the model Hamiltonian (\ref{eq:H}), below we obtain the memory kernel of the
NIBA equation exactly, beyond the so-called ``linear response" approximation. This form should be adopted when
the thermal ``bath" is finite (the case in realistic applications), the spectral density function is non-continuous,
 and strongly-coupled discrete modes are prominent. Then, the mapping to the spin-boson model does not hold \cite{MakriSpin}
and deviations may be significant, particularly in the scaling limit $\omega_c/\Delta \gg1$ \cite{Wang}.
Our expression for the kernel reduces to the ``thermodynamic" limit discussed
in the literature \cite{Silbey,Makriproof} when the interaction with the subsystem is well diluted over many spins.

From the technical point of view,
in the SB model the Campbell-Baker-Hausdorff formula can be applied in its simple form
$e^{cb^{\dagger}-c^*b}=e^{cb^{\dagger}}e^{-c^*b}e^{-|c^2|/2 }$, $c$ is a c-number
and $b$ a bosonic operator. This relation holds since $[b,b^{\dagger}]=1$,
rendering an exponential form for the memory kernel.
Such a relation does not hold for the SS model, resulting in a non-exponential form. 

The derivation of NIBA equations begins with the Liouville equation for the total density matrix $\rho$.
Compacting the Hamiltonian into $\tilde H_{SS} = \tilde H_0+\tilde V$, with $\tilde V$ comprising the $\Delta$-product term
in Eq. (\ref{eq:Htilde}), we write
\bea
\dot\rho(t)=-i[\tilde V(t),\rho(0)]-\int_0^t d\tau [\tilde V(t),[\tilde V(\tau),\rho(\tau]].
\label{eq:EOM1}
\eea
Here $\tilde V(t)$ is an operator in the interaction representation.  
As an initial condition we take a factorized form $\rho(0)=\rho_S(0)\rho_B$ with the subsystem reduced density matrix
$\rho_S(0)$ and the two-bath factorized state $\rho_B=\rho_{L}\rho_R$,
$\rho_{\nu}=e^{-\beta_{\nu}H_{\nu}}/Z_{\nu}$
with $H_{\nu}=\frac{1}{2}\sum_{j}\omega_{j}\sigma_z^{\nu,j}$. Here $Z_{\nu}$ is the partition function of the $\nu$ bath.
Note that the initial state assumes the unperturbed form.
We now trace over the baths' degrees of freedom and make the approximations
that their operators can be decoupled from the subsystem \cite{Dekker}, and that they maintain the initial state,
\bea
{\rm tr_B}[{e^{i\Omega(t)}e^{-i\Omega(0)}\rho(t)}]\sim {\rm tr_B}[{e^{i\Omega(t)}e^{-i\Omega(0)}\rho_B}]\rho_S(t).
\eea
These assumptions can be justified in different limits \cite{Legget}, particularly at high temperatures and
when the coupling to the bath is not extremely strong, so as system-bath correlations can be ignored.
If we assume a diagonal initial condition for the subsystem, the first term in Eq. (\ref{eq:EOM1}) drops,
and we reach the following equation of motion for the population of the subsystem, 
$p_1(t)-p_0(t)={\rm tr}[\sigma_z \rho(t)]$,
\bea
\frac{dp_1(t)}{dt}&=& -\frac{\Delta^2}{4}
\int_{0}^{t}d\tau
\left[ e^{i\omega_0(t-\tau)} \langle e^{-i\Omega(t)}e^{i\Omega(\tau)}\rangle +
e^{-i\omega_0(t-\tau)} \langle e^{-i\Omega(\tau)}e^{i\Omega(t)} \rangle
\right] p_1(\tau)
\nonumber\\
& + &  \frac{\Delta^2}{4}
\int_{0}^{t} d\tau
\left[ e^{i\omega_0(t-\tau)} \langle e^{i\Omega(\tau)}e^{-i\Omega(t)}\rangle  +
e^{-i\omega_0(t-\tau)} \langle e^{i\Omega(t)}e^{-i\Omega(\tau)}\rangle
\right]p_0(\tau),
\nonumber\\
1&=&p_0(t)+p_1(t).
\label{eq:P1}
 \eea
Next, we define the correlation function
\bea
e^{-Q_S(t)}\equiv\langle e^{i\Omega(t)}e^{-i\Omega(0)}
\rangle,
\label{eq:corr}
\eea
with the thermal average performed over the $L$ and $R$ reservoirs' degrees of freedom.
Below we explicitly confirm that $\langle e^{i\Omega_j(t)} e^{-i\Omega_j(0)}\rangle$ = $\langle e^{-i\Omega_j(t)} e^{i\Omega_j(0)}\rangle$, thus we can re-write the NIBA equation in the more common form as
\bea \frac{dp_1(t)}{dt}&=& -\frac{\Delta^2}{2} \int_{0}^{t}
e^{-Q_S'(t-\tau)} \cos[ \omega_0(t-\tau)-Q_S''(t-\tau)] p_1(s) d\tau
\nonumber\\
& +&\frac{\Delta^2}{2} \int_{0}^{t} e^{-Q_S'(t-\tau)} \cos[
\omega_0(t-\tau)+Q_S''(t-\tau)] p_0(\tau)d\tau.
\label{eq:P2}
 \eea
The function $Q_S(t)=Q_S'(t)+iQ_S''(t)$ includes a real and an imaginary component.

\subsection{Memory kernel}
\label{kernel}

In this section we evaluate the correlation function (\ref{eq:corr}).
We drop the $\nu$ index in steps when it is inconsequential;
it re-appears in final expressions when spins in both baths should be clearly accounted for.
We begin with a mode $j$ of frequency $\tilde \omega_j$, 
use the identity 
$e^{ia\sigma_z}=\cos(a)+i\sin(a)\sigma_z$, and write
\bea
i\Omega_j(t)&\equiv& i \arctan(\eta_j)\sigma_y^{j}(t)
\nonumber\\
&=&  i \arctan(\eta_j) \left[\cos (\tilde\omega_jt)\sigma_y^j +\sin(\tilde\omega_j t)\sigma_x^j\right].
\eea
This operator is exponentiated by making use of the relation 
\bea
e^{i(a_y\sigma_y + a_x\sigma_x)}&=&
\cos\sqrt{a_x^2+a_y^2}
\nonumber\\
 &+&i\sin\sqrt{a_x^2+a_y^2}
\frac{a_y\sigma_y+a_x\sigma_x}{\sqrt{a_x^2+a_y^2}},
\eea
providing,
\bea
e^{i\Omega_j(t)}=
\frac{1}{(1+\eta_j^2)^{1/2}} + i \frac{\eta_j}{(1+\eta_j^2)^{1/2}}\left[
\cos(\tilde\omega_j t)\sigma_y^j + \sin(\tilde\omega_j t) \sigma_x^j\right].
\eea
We now consider the product
\bea
e^{i\Omega_j(t)} e^{-i\Omega_j(0)}&=&
\frac{1}{1+\eta_j^2}
\left\{ 1 + i\eta_j
\left[\cos(\tilde\omega_j t)\sigma_y^j +\sin(\tilde\omega_j t)\sigma_x^j \right]\right\}
\nonumber\\
&\times&
(1-i\eta_j\sigma_y^j),
\eea
and trace the bath mode (spin $j$) with the corresponding thermal weight,
\bea
&&{\rm tr}_j
[e^{-\beta\omega_j\sigma_z^j/2} e^{i\Omega_j(t)} e^{-i\Omega_j(0)} ]/Z_j
\nonumber\\
&&= \frac{1}{1+\eta_j^2}
\left[  1+\eta_j^2\cos(\tilde\omega_j t) -i\eta_j^2\sin(\tilde\omega_j t)
\tanh\left(\frac{\beta\omega_j}{2} \right) \right].
\eea
Here the partition function for the $j$ spin is
\bea
Z_j=e^{-\beta\omega_j/2}+e^{\beta\omega_j/2},
\eea
and the temperature corresponds to the particular bath temperature, $j\in \nu$.
We recall that $\eta_j=\lambda_j/\omega_j$,
and write the correlation function as
\bea
\langle e^{i\Omega_j(t)} e^{-i\Omega_j(0)}\rangle =
\frac{1}{1+\lambda_j^2/\omega_j^2}
\left[ 1 + \frac{\lambda_j^2}{\omega_j^2}\cos(\tilde\omega_j t) -i\frac{\lambda_j^2}{\omega_j^2} \sin(\tilde \omega_j t)
\tanh \left( \frac{\beta\omega_j}{2}\right)  \right].
\label{eq:R2}
\eea
It is easy to confirm that $\langle e^{i\Omega_j(t)} e^{-i\Omega_j(0)}\rangle$ = $\langle e^{-i\Omega_j(t)} e^{i\Omega_j(0)}\rangle$.
This procedure is repeated - independently- for each spin within the $L$ and $R$ reservoirs. The total-discretized spin bath
($d-S$) correlation function is given in a product form
\bea
C_{d-S}(t)&\equiv&\langle e^{i\Omega(t)} e^{-i\Omega(0)}\rangle
\nonumber\\
&=&
\Pi_{j\in L} \langle e^{i\Omega_j(t)} e^{-i\Omega_j(0)}\rangle
\Pi_{j\in R} \langle e^{i\Omega_j(t)} e^{-i\Omega_j(0)}\rangle.
\label{eq:R3}
\eea
Equations (\ref{eq:R2})-(\ref{eq:R3}) are exact and meaningful as a memory kernel even if the reservoirs
include discrete components, as long as the resulting dynamics is consistent with the basic approximations:
the factorization of the correlation function into a subsystem-bath product form, and the assumption that the reservoirs are maintained
(each) in a thermal equilibrium state.
It is not difficult to confirm that Eqs. (\ref{eq:R2})-(\ref{eq:R3})  reduce to the so-called ``linear response" result in the
thermodynamic limit when the spectral density function is continuous and subsystem-bath interactions are
well diluted over many spins \cite{Makriproof,MakriSpin, Wang}.
We show this by (ignoring the trivial $\nu$-bath identifier)
assuming that $\lambda_j/\omega_j\ll1$ for all spins, see 
explanations below Eq. (\ref{eq:Jeff}).
We can now expand the denominator of Eq. (\ref{eq:R2}), $1/(1+\eta_j^2)\sim (1-\eta_j^2)$, and
collect lowest-order terms in $\lambda_j^2/\omega_j^2$. This leads to
\bea
&&C_S(t)\equiv
\langle e^{i\Omega(t)} e^{-i\Omega(0)}\rangle_{\eta_{j}\ll1}
\nonumber\\
&&\sim \Pi_{j\in \nu}
\left[1-\frac{\lambda_j^2}{\omega_j^2}
\left(1-\cos(\omega_j t)\right)
-i\frac{\lambda_j^2}{\omega_j^2}\sin(\omega_j t) \tanh \left( \frac{\beta_{\nu}\omega_j }{2}\right) \right].
\label{eq:app1}
\eea
We approximate this result with an exponential form,  and reach (recovering the $\nu$ marker),
\bea
C_S(t)
&\sim&
e^{-\sum_{\nu, j}\frac{\lambda_{\nu,j}^2}{\omega_j^2}
\left[ (1- \cos(\omega_j t))  + i \sin(\omega_j t) \tanh \left( \frac{\beta_{\nu}\omega_j}{2} \right) \right] }
\nonumber\\
&=&C_{S,L}(t)C_{S,R}(t)
\label{eq:app}
\eea
If the frequencies are distributed continuously, we can define the
spectral density function, possibly distinct at the two terminals,
\bea
J_{\nu}(\omega)=\pi\sum_{j}|\lambda_{\nu,j}|^2\delta(\omega-\omega_j).
\label{eq:J1}
\eea
Specifically, we use in the simulations of Sec. \ref{Transport} an Ohmic form
\bea
J_{\nu}(\omega)=2\pi\alpha_{\nu}\omega e^{-\omega/\omega_c},
\label{eq:J2}
\eea
where the dimensionless Kondo parameter $\alpha_{\nu}$ measures the strength of the subsystem-bath interaction.
We are interested in both a spatially symmetric junction with $\alpha_L=\alpha_R$, and an asymmetric setup with
$\alpha_L>\alpha_R$. We work in the non-adiabatic limit, thus the cutoff frequency $\omega_c$ is
taken large, $\Delta/\omega_c\ll1$. For simplicity, $\omega_c$ is set identical for the two baths.

Using the definition (\ref{eq:J1}), the exponent of the correlation function
$C_{S,\nu}(t)=e^{-Q_{S,\nu}(t)}$,  $Q_{S,\nu}(t)=Q_{S,\nu}'(t)+iQ_{S,\nu}''(t)$,
is given by
%
\bea
Q_{S,\nu}'(t)&=&\int_0^{\infty}d\omega \frac{J_{\nu}(\omega)}{\pi\omega^2}[1-\cos(\omega t)],
\nonumber\\
Q_{S,\nu}''(t)&=&
\int_0^{\infty}d\omega \frac{J_{\nu}(\omega)}{\pi\omega^2}
\sin(\omega t) \tanh \left( \frac{\beta_{\nu}\omega}{2} \right).
\label{eq:QS}
\eea
This result  has been introduced in Refs. \cite{CaldeiraSpin,Makriproof}, and it corresponds to the
correlation function of the harmonic bath [see Eq. (\ref{eq:QH})] once we define a temperature dependent,
effective, spectral density function as
\bea
J^{eff}_{\nu}(\omega, \beta_{\nu}) = \tanh \left( \frac{\beta_{\nu}\omega}{2}\right) J_{\nu}(\omega).
\label{eq:Jeff}
\eea

We now discuss the assumptions behind the linear-response ``thermodynamic"  limit by following Ref. \cite{Makriproof}.
The spectral density function relates to the density of states
$\rho(\omega)=\sum_j\delta(\omega-\omega_j)$ via
\bea
\pi \lambda_j^2 \sim J(\omega_j) /\rho(\omega_j).
\label{eq:gj}
\eea
Let us assume that the spin bath has a continuous spectra, for example,
$\rho(\omega) \sim \frac{N_B}{\omega_c}e^{-\omega/\omega_c}$ with 
$N_B$ the number of spins in the bath. This function corresponds to the (Ohmic) spectral density function, 
$J(\omega)\sim \rho(\omega)\omega$. Using these expressions in Eq. (\ref{eq:gj}) we find that $\lambda_j^2\sim \alpha
\frac{\omega_j\omega_c}{N_B}$ with $\alpha$ as the dimensionless Kondo parameter. In the thermodynamic
limit, $N_B\rightarrow\infty$, thus $\lambda_j^2/\omega_j^2\rightarrow 0$. Therefore, as long as the density of
states can be represented by a continuous-dense function with uniform couplings to the subsystem,
the assumption $\lambda_j^2/\omega_j^2\ll1$ holds in the thermodynamic limit, and the harmonic-like form (\ref{eq:app}) applies.
However, local modes: defects, and impurity spins, typically
couple to the central spin via an interaction energy $\lambda_j$ which is {\it independent} of $N_B$ \cite{Stamp}.
In such cases the discrete form  (\ref{eq:R2})-(\ref{eq:R3}) should be employed.
In general situations 
one could facilitate the numerics by constructing NIBA's kernel as a product of Eq. (\ref{eq:R2}), including all
strongly-coupled modes, and
Eq. (\ref{eq:app}), containing modes that are weakly coupled to the subsystem.

\begin{figure}[htbp]
\vspace{0mm} \hspace{0mm} {\hbox{\epsfxsize=85mm \epsffile{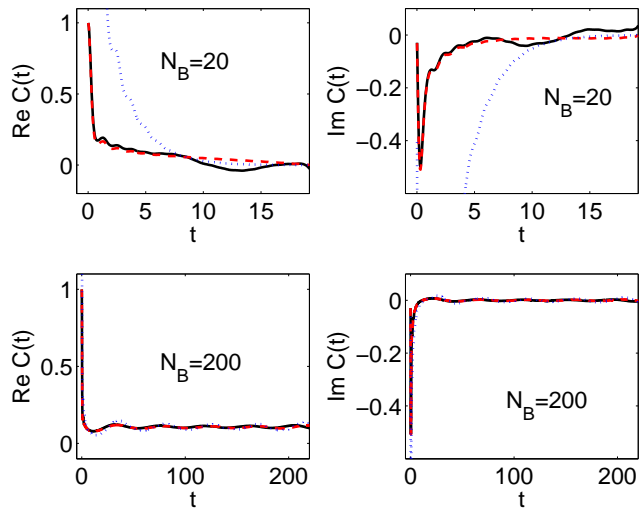}}}
\caption{Bath correlation function (\ref{eq:corr})
for a spin bath, using the exact form (\ref{eq:R2})-(\ref{eq:R3}) (full),
the approximate form (\ref{eq:app}),
valid in the thermodynamic limit (dashed), and the
harmonic-bath result (\ref{eq:CH}) (dotted).
Parameters are $\beta$=5, $\alpha=0.4$, a lower frequency cutoff for the bath $\omega_i=0.15$,
and an upper cutoff $\omega_c=5$.}
\label{Fig1}
\end{figure}

In Fig. \ref{Fig1} we display the correlation function (\ref{eq:corr}) using the exact form  $C_{d-S}(t)$
and the thermodynamic limit $C_S(t)$. We use a bath with a constant density of states,
$\rho=N_B/(\omega_c-\omega_i)$; $\omega_{i}$ and $\omega_{c}$ are low and high energy cutoffs, respectively.
We also display the correlation function (\ref{eq:CH}) in the case of a harmonic bath,
and show that harmonic baths and spin baths similarly behave in the thermodynamic limit at low temperatures.
Results are displayed up to the recurrence time $\tau_{rec}\sim 2\pi/\rho$. Deviations between the discrete
and continuous forms are observed when the bath is small, $N_B=20$, since in this case $\eta_j\sim 1$ for low frequencies.

The refined kernel can be employed for the calculation of transfer rates in anharmonic media, in the non-adiabatic regime.
In the context of donor-acceptor electron transfer processes in condensed phases,
one can define a meaningful transfer rate
if the donor population exhibits an exponential decay.
This is indeed the case when the bath decorrelation time is short relative to the electron
dynamics, when the bath temperature is sufficiently high.
In this case, limiting the discussion to a single spin bath,
the donor population follows a kinetic equation $\dot p_1=-k_d(t)p_1(t) +k_u(t)p_0(t)$
with the time-dependent rate
%
\bea
k_d(t) =  \frac{\Delta^2}{2}
\int_{0}^t d\tau
e^{-Q'_S(\tau)} \cos[\omega_0\tau - Q''_S(\tau)]d\tau.
\label{eq:rateC}
\eea
This definition is meaningful if $k_d(t)$ reaches a constant value after a certain time (shorter than the
subsystem's characteristic time).
The long-time limit, a standard Fermi-golden-rule rate, is given by
\bea
k_d =  \frac{\Delta^2}{4}
\int_{-\infty}^\infty d\tau
e^{i\omega_0 \tau}e^{-Q_S(\tau)} d\tau.
\label{eq:rateF}
\eea
In the next section we explain the application of the NIBA equations to problems of heat transport through a TLS
interfacing two separate (spatially or spectrally) thermal baths.

%


\section{Heat Transport}
\label{Transport}

The ability to control and manipulate energy flow at the nanoscale is critical in different technologies.
In molecular electronics one needs to
effectively remove the dissipated heat from the active area to ensure molecular stability \cite{MolEl}.
Phonon engineering is a key challenge in applications involving thermal management, in energy conversion devices \cite{Balandin},
and in small-scale thermometry and refrigeration systems \cite{Pekola}.
More recently, ''phononic"  devices have been proposed, for mimicking the (traditional) 
functionality of electronic systems, transistors and memory units \cite{Phononics}.
Specific elements of recent interest are thermal diodes,
rectifying thermal energy current, and junctions showing negative differential thermal conductances \cite{Phononics}.

A thermal diode is an elementary two-terminal device. It transports heat more effectively
in one direction of the temperature bias than in the reversed direction \cite{RectifReview}.
This effect has been investigated in detail experimentally and theoretically 
at the macro-micro scale \cite{macro,SolidD}, the nanoscale \cite{RectifE}, and more recently in the molecular realm \cite{Dlott1, Dlott2,Leitner}, with
proposals relying on phononic, electronic, and photonic heat conduction \cite{RectifReview}.
It was also recently demonstrated that Josephson tunnel junctions (Cooper-pair condensates electrodes)
can act as strong phase-tunable thermal diodes \cite{Giazotto}.

The question of the emergence of the diode effect and NDTC,
when different interfaces are employed, has been explored in numerous studies, see for example
Refs. \cite{Casati0,Casati,SegalRectif,SegalM,Phononics,NDTC,BaowenI,Ren}.
In what follows we work at the level of NIBA and demonstrate that
both the SS and SB models can support the diode behavior, yet significant differences in the transport behavior
manifest themselves, particularly in the weak-to-intermediate coupling regime.
The SS and SB models also acquire distinct NDTC behavior, see the discussion below Eq. (\ref{eq:gaussJS2}).

\subsection{NIBA: Heat current expression}

We consider a TLS between two thermal baths comprising either non-interacting spins or
bosons, and provide a closed expression for the steady-state energy (in the form of heat) current in this system.
We explore the heat transport characteristics under the NIBA as discussed in Sec. \ref{Model}. We note that
the equation of motion  (\ref{eq:P2}) holds for a TLS coupled to spin baths, with the kernel
$C_{d-S}(t)$ or $C_S(t)$, or to harmonic baths, with the memory
function $C_H(t)$, see Appendix A. To unify our presentation, we refer below to the
correlation function in Eq. (\ref{eq:P1}) as $C(t)$, corresponding to either cases.

Our discussion follows Refs. \cite{SegalRectif,SegalM,FR2}.
In the Markovian limit the subsystem evolves slowly in comparison to the reservoirs dynamics. We then make two simplifications to the integro-differential equation (\ref{eq:P2}): 
We replace the population, $p_n(\tau)$ by the time-local value
$p_n(t)$ ($n=0,1$). This approximation is justified when the timescale over which the memory (represented by
the integral) is significant, is short in comparison to the time interval for significant changes in the subsystem population. We also extend the limits of the integrals to infinity, again relying on the argument of time-scale separation, with the integrand quickly dying out. With that, Eq. (\ref{eq:P1}) reduces to the kinetic form
\bea \dot p_1&=&-k_dp_1(t) +k_up_0(t).
\label{eq:EOMm} \eea
The rate constants are given by Fourier transforms of bath correlation functions,
\bea
k_d &=&
C(\omega_0),\,\,\,\,\,\,
k_u =
C(-\omega_0),
\nonumber\\
C(\omega_0) &=&  
\int_{-\infty}^\infty
e^{i\omega_0 t} C_{L}(t) C_{R}(t) dt.
\label{eq:Cw0}
\eea
Using the convolution theorem, the transition rates can be  written as a convolution of reservoirs-induced processes,
\bea C(\omega_0)&  = & \frac{1}{2\pi}\int_{-\infty}^\infty
C_{L}(\omega_0 - \omega)C_{R}(\omega) d\omega, \label{eq:conv} \eea
with the Fourier transform
\bea
C_{\nu}(\omega)=\int_{-\infty}^{\infty}e^{i\omega t}C_{\nu}(t)dt.
\label{eq:Cnu}
\eea
This rate satisfies the detailed balance relation,
\bea
\frac{C_{\nu}(\omega)}{C_{\nu}(-\omega)}&  = & e^{\beta_{\nu}\omega}.
\label{eq:DB}
\eea
This relation does not hold for the combined rate $C(\omega)$, as it allows for non-additive processes: When
the subsystem decays it disposes an energy $\omega_{0}$ into both reservoirs: the amount $\omega$ is
dissipated into the $R$ bath while the $L$ bath gains (or contributes) the rest $\omega_{0}-\omega$.
Similarly, excitation of the subsystem involves both reservoirs in a non-additive manner. Since energy is
dissipated or absorbed in compound processes, counting the number of transitions between subsystem
levels within a certain time interval does not correspond to the amount of energy transferred between the
reservoirs during that interval. The (nontrivial) resolved master equation, discussed in Ref. \cite{FR2}
for the SB model, holds for the SS model as well.
We then derive the cumulant generating function for the SS model,
confirm the fluctuation theorem, and obtain a closed expression for the steady-state energy current, defined
positive when flowing left to right,
\bea \langle J \rangle=
\left(\frac{\Delta}{2}\right)^2 \frac{1}{2\pi}\int_{-\infty}^{\infty}\omega d\omega
\left[ C_R(\omega)C_L(\omega_0-\omega)p_1 -
C_R(-\omega)C_L(-\omega_0+\omega)p_0 \right].
\label{eq:Current} \eea
Here, the population of the TLS corresponds to the steady-state limit ($\dot p=0$) with
\bea
p_1=C(-\omega_0)/[C(\omega_0)+C(-\omega_0)], \,\,\,\,
p_0=C(\omega_0)/[C(\omega_0)+C(-\omega_0)]. \label{eq:SSpop}
\eea
Since we set $\hbar=1$, the energy current has the dimension $Energy^2$.
In the first paragraph of Sec. \ref{simulation} we convert
the current to physical units.

Concluding, Eq. (\ref{eq:Current})
is valid in the non-adiabatic $\Delta/\omega_c \ll1$
limit. If the reservoirs include non-interacting spins 
the correlation function  $C_{S,\nu}(t)$  [or $C_{d-S}(t)$ and its $\nu$ components]
should be used. In the case of harmonic baths,
$C_{H,\nu}(t)$ from Eq. (\ref{eq:CH}) should be adopted.
One can also consider a composite case with the subsystem mediating harmonic and anharmonic (spin) reservoirs,
or when each reservoir includes both normal modes and localized spin modes.
In the low-temperature limit
$\tanh(x)\xrightarrow{x\rightarrow\infty}1$, $Q_S(t)$ and $Q_H(t)$ come together,
and the thermal conductances coincide. Analytic results for the high-temperature regime are presented in the next subsection.

\subsection{NIBA: Analytic results}
\label{Analytics}

We derive analytic expressions for the heat current in the SB and the SS models 
by extending previous studies \cite{SegalRectif,SegalM,FR2}. The results
serve to pinpoint the fundamentally different transport characteristics of these two models.
Below we denote by $\delta T=T_L-T_R$ the applied temperature bias. The average temperature is referred to as
$T_a=(T_L+T_R)/2$.
We begin our discussion with the SB model, summarizing results from Ref. \cite{SegalRectif}.
Assuming sufficiently high temperatures and strong coupling, we perform a short-time expansion of $Q_{H,\nu}(t)$
[Eq. (\ref{eq:QH})] to reach
\bea
Q'_{H,\nu}(t)&=&T_{\nu}t^2\int_{0}^{\infty}d\omega \frac{J_{\nu}(\omega)}{\pi \omega} = E_r^{\nu}T_{\nu}t^2,
\nonumber\\
Q''_{H,\nu}(t)&=&t\int_{0}^{\infty}d\omega \frac{J_{\nu}(\omega)}{\pi \omega}=E_r^{\nu}t.
\label{eq:QHshort}
\eea
Here, we define the reorganization energy as
$E_r^{\nu}=\int_{0}^{\infty}d\omega \frac{J_{\nu}(\omega)}{\pi \omega}$.
With this Gaussian, Marcus-type, kernel, we can obtain a closed expression for the heat current (\ref{eq:Current}),
\bea \langle J_H\rangle = \Delta^2\frac{\sqrt{2\pi}E_r^LE_r^R\delta T}{(2E_r^L
T_L+2E_r^R T_R)^{\frac{3}{2}}} \exp\left[-\frac{(E_r^L + E_r^R -
\omega_0)^{2} }{4(E_r^L T_L+E_r^R T_R)}\right]\times f_H
\label{eq:gaussJ}
\eea
where $f_H = \left\{ \exp\left[\frac{\omega_0(E_r^L + E_r^R)}{(E_r^L
T_L+E_r^R T_R)}\right] + 1\right\}^{-1}$.
We clear-up this expression by using 
the Ohmic spectral function  Eq. (\ref{eq:J2}), to provide $E_r^{\nu}=2\alpha_{\nu}\omega_c$,
 then setting $\omega_0=0$. Eq. (\ref{eq:gaussJ}) now takes the more-transparent structure
\bea \langle J_H\rangle = \frac{\Delta^2}{4}\frac{\sqrt{2\pi\omega_c}\alpha_L\alpha_R\delta T}{(\alpha_L
T_L+\alpha_R T_R)^{\frac{3}{2}}} \exp\left[-\frac{\omega_c(\alpha_L+\alpha_R)^{2} }{2(\alpha_L T_L+\alpha_R T_R)}\right].
\label{eq:gaussJ2}
\eea
This expression reveals the effect of thermal rectification as discussed in Refs. \cite{SegalRectif,SegalM}.
However, while in general the SB model can support NDTC, see Eq. (\ref{eq:gaussJ}),
in symmetric setups, $\alpha_L=\alpha_R$, and for an unbiased TLS,
$\omega_0=0$, the current is strictly linear with $\langle J_H \rangle \propto \delta T$.

We now turn to the spin-spin-bath model and again pursue the tractable limit of a Gaussian kernel.
Assuming that the time correlation function (\ref{eq:corr}) quickly decays (in comparison to the subsystem's dynamics),
we perform a short-time expansion of Eq. (\ref{eq:QS}). Adopting the Ohmic function for the spectral density we reach
\bea
Q'_{S,\nu}(t)&=&t^2\int_{0}^{\infty}d\omega \frac{J_{\nu}(\omega)}{2\pi} = \alpha_{\nu}\omega_c^2t^2,
\nonumber\\
Q''_{S,\nu}(t)&=&\frac{t}{2\pi T_{\nu}}\int_{0}^{\infty}d\omega J_{\nu}(\omega)=\frac{\alpha_{\nu}\omega_c^2}{T_{\nu}}t.
\label{eq:QSshort}
\eea
The real part is ruled by the cutoff frequency, and it is insensitive to the temperature which is now controlling
the period of oscillation,  shorter at low temperatures. Consistently with NIBA assumptions, the memory function 
strongly decays in the non-adiabatic limit, $\omega_c\gg \Delta$.
If we define an effective reorganization energy
\bea
\tilde E_r^{\nu}\equiv \frac{\alpha_{\nu}\omega_c^2}{T_{\nu}},
\label{eq:reorgSS}
\eea
we immediately restore the harmonic $Q_{H,\nu}(t)$ [Eq. (\ref{eq:QHshort})]. As a result, the heat current
in the SS model follows equation (\ref{eq:gaussJ}), only with $\tilde E_r^{\nu}$ replacing $E_r^{\nu}$.
Using again the Ohmic form for the spectral density, the current in the SS model overall obeys
\beq \langle J_S\rangle = \Delta^2\frac{\sqrt{\pi}\alpha_L\alpha_R\omega_c}{2(\alpha_L
+\alpha_R)^{\frac{3}{2}}}\frac{\delta T}{T_L T_R} \exp\left\{-\frac{[\omega_c^2(\frac{\alpha_L}{T_L}+\frac{\alpha_R}{T_R}) -
\omega_0]^{2} }{4(\alpha_L+\alpha_R)\omega_c^2}\right\}\times f_S,
\label{eq:gaussJS}
\eeq
where $f_S = \left\{ \exp\left[\frac{\omega_0(\frac{\alpha_L}{T_L} +\frac{\alpha_R}{T_R} )}{(\alpha_L+\alpha_R)}\right] + 1\right\}^{-1}$.
In the unbiased case it becomes
\beq \langle J_S\rangle = \frac{\Delta^2}{4}\frac{\sqrt{\pi}\alpha_L\alpha_R\omega_c}{(\alpha_L
+\alpha_R)^{\frac{3}{2}}} \frac{\delta T}{T_L T_R}\exp\left[-\frac{\omega_c^2(\frac{\alpha_L}{T_L}+\frac{\alpha_R}{T_R})^{2} }{4(\alpha_L+\alpha_R)}\right].
\label{eq:gaussJS2}
\eeq
For simplicity, we compare results in the unbiased limit, Eq. (\ref{eq:gaussJ2}) 
and Eq. (\ref{eq:gaussJS2}).
Both expressions scale with the subsystem-bath coupling as $\langle J \rangle \propto \sqrt \alpha e^{-\epsilon\alpha}$
where $\epsilon\propto(\omega_c/T_a)^n$; $n=1$ ($n=2$) for the SB (SS) model.
Furthermore, in both junctions the current shows a crossover from a weak-coupling regime 
(see a careful discussion below),
in which the current increases as $\sqrt{\alpha}$, to the strong-coupling limit,
when the current exponentially decays with $\alpha$.

Eqs. (\ref{eq:gaussJ2}) and (\ref{eq:gaussJS2}) uncover
significant qualitative differences between the models, concerning NDTC and the thermal diode effect: (i)
For relatively small $\alpha$ the transport behavior is dominated by the prefactor of the exponent. In the case of an SB junction
this prefactor allows for thermal rectification when $\alpha_L\neq\alpha_R$; the combination
$(\alpha_LT_L+\alpha_RT_R)$ is sensitive to the direction of the applied temperature bias. In contrast,
the SS model cannot exhibit thermal rectification in the small-$\alpha$
limit since the prefactor in Eq. (\ref{eq:gaussJS2}) is left unaffected under the exchange of temperatures.
(ii) In the SS model the current is higher when the hot contact is strongly coupled to the subsystem,
than the opposite case, when the hot contact is weakly coupled to the subsystem,
for $\alpha_L>\alpha_R$ and positive $\delta T$, $|J(\delta T)|>|J(-\delta T)|$.
In contrast, the SB model manifests an involved behavior: For small $\alpha$
the prefactor dominates, and the current is larger when the hot terminal is weakly coupled to the subsystem.
Only at very large $\alpha$, when the exponential factor dominates the current, this trend is reversed.
(iii) NDTC is missing in the SB model if the setup is symmetric $\alpha_L=\alpha_R$ and unbiased
$\omega_0=0$, see Eq. (\ref{eq:gaussJ2}). This observation stands in a sharp contrast to the SS junction,
which under the same conditions follows $\langle J_S\rangle \propto \frac{\delta T}{T_a^2-\delta T^2/4}
\exp\left[-\frac{\alpha\omega_c^2 T_a}{2(T_a^2-\delta T^2/4)}\right]$. 
The SS model thus supports a significant NDTC already
at intermediate coupling. In numbers,
when $\alpha=0.4$, $T_a=2\Delta$ and $\omega_c=10\Delta$, the current decreases by an order of magnitude when
$\delta T/\Delta$ is increased from 1 to 2.

\begin{figure}[htbp]
\vspace{0mm} \hspace{0mm}
{\hbox{\epsfxsize=75mm \epsffile{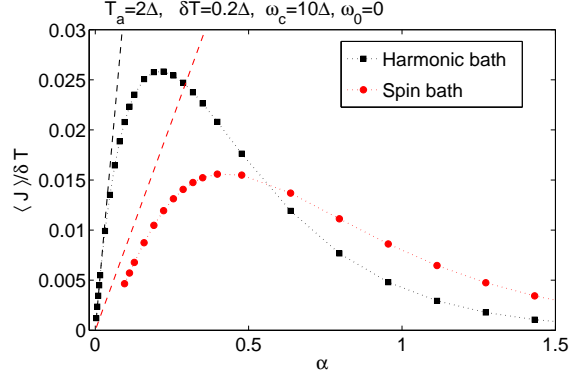}}}
\caption{
Energy current in the SB  ($\square$) and the SS  ($\circ$) models,
using NIBA and the weak-coupling limit, Eqs. (\ref{eq:weakS}) and (\ref{eq:weakH})
(dashed lines).
Parameters appear in the figure, $\alpha=\alpha_{L,R}$.}
\label{FigJ2}
\end{figure}

We clarify now on the range of the ``weak-coupling regime" in the present  discussion. 
To be consistent with the short-time expansion of the kernel, the bath relaxation time should be made short.
In the case of a spin bath, Eq. (\ref{eq:QSshort}) provides 
$1/\sqrt{\alpha \omega_c^2}\ll\Delta^{-1}$, or $(\Delta/\omega_c)^2\ll \alpha$. 
For $\Delta/\omega_c=0.1$,
coupling strengths of $\alpha>0.1$ are consistent with the underlying assumptions.
On the other hand, if we wish the prefactor in Eq. (\ref{eq:gaussJS2}) to dominate the current
we should consider a small exponent, $\alpha(\omega_c/T_a)^2/2<1$. If $T_a/\omega_c=0.2$,
we are bound from above by $\alpha<0.2$.
Thus, the weak-coupling limit of NIBA corresponds to the relatively narrow regime of $0.1<\alpha<0.2$, see simulations
in Fig. \ref{FigJ2}.


\begin{figure}[htbp]
\vspace{0mm} \hspace{0mm}
{\hbox{\epsfxsize=75mm \epsffile{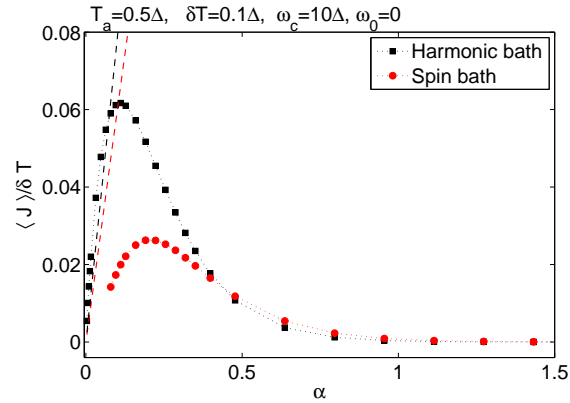}}}
\caption{
Low-temperature energy current in the SB model  ($\square$) and the SS case ($\circ$)
from NIBA and the weak-coupling limit (dashed lines).
Parameters appear in the figure, $\alpha=\alpha_{L,R}$.%
}
\label{FigJ2low}
\end{figure}

\begin{figure}[htbp]
\vspace{0mm} \hspace{0mm}
{\hbox{\epsfxsize=70mm \epsffile{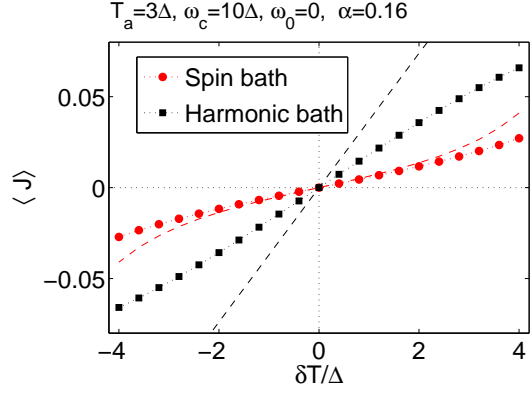}}}
\caption{
Energy current in the SB ($\square$) and the SS ($\circ$) models
as a function of $\delta T/\Delta$.
The dashed lines correspond to the Born-Markov limit of the SS 
(\ref{eq:weakS}) and the SB models (\ref{eq:weakH}).
Parameters appear in the figure.
}
\label{Figd0}
\end{figure}


\subsection{Born-Markov expressions}

In the strict weak-coupling limit ($\alpha\ll1$) assuming wide-band reservoirs,  $\Delta\ll\omega_c$,
one can use the Born-Markov approximation and derive analytical expressions for the
energy current in the unbiased ($\omega_0=0$) SS and SB models \cite{Claire}.
In the SS model the current follows
\bea
\langle J_{S,w}\rangle=\Delta \frac{\Gamma_L(\Delta)\Gamma_R(\Delta)}{\Gamma_L(\Delta)+
\Gamma_R(\Delta)}\left[ n_S^L(\Delta)-n_S^R(\Delta) \right],
\label{eq:weakS}
\eea
with the spin occupation factor
\bea
n_S^{\nu}(\Delta)=[e^{\beta_{\nu}\Delta}+1]^{-1}.
\eea
The coupling energy
$\Gamma_{\nu}(\Delta)=\frac{\pi}{2}\sum_{j}\lambda_{\nu,j}^2\delta(\Delta-\omega_j)$ 
is closely related to the spectral density function
\cite{Claire,FR2}.
We now define the Bose-Einstein distribution function
\bea
n_B^{\nu}(\Delta)=[e^{\beta_{\nu}\Delta}-1]^{-1},
\eea
and recall the identities
$\coth(x/2)=2n_B(x)+1$, $n_B(x)=[2n_B(x)+1]n_S(x)$. We also
define an effective coupling energy as
\bea
\tilde \Gamma_{\nu}(\Delta)\equiv\Gamma_{\nu}(\Delta)\tanh(\beta_{\nu}\Delta/2).
\eea
The energy current in the SS model, Eq. (\ref{eq:weakS}), now reduces to
\bea
\langle J_{S,w}\rangle=
\Delta \frac{\tilde\Gamma_L(\Delta)\tilde\Gamma_R(\Delta)}
{[1+2n_B^L(\Delta)]\tilde\Gamma_L(\Delta)+[1+2n_B^R(\Delta)]
\tilde\Gamma_R(\Delta)}[n_B^L(\Delta)-n_B^R(\Delta)].
\label{eq:weakSH}
\eea
This expression corresponds to the heat current in the SB model \cite{Claire, FR2} once restoring
the temperature independent coupling $\tilde \Gamma\rightarrow \Gamma$,
\bea
\langle J_{H,w}\rangle=\Delta \frac{\Gamma_L(\Delta)\Gamma_R(\Delta)}{[1+2n_B^L(\Delta)]\Gamma_L(\Delta)
+[1+2n_B^R(\Delta)]
\Gamma_R(\Delta)}[n_B^L(\Delta)-n_B^R(\Delta)].
\label{eq:weakH}
\eea
Hence, we have confirmed that the mapping of the spectral density function (\ref{eq:Jeff}) correctly
converts the SS heat current to the SB case.

At low temperatures, $\beta_{\nu}\Delta\gg1$, $n_{S,B}^{\nu}(\Delta)\rightarrow e^{-\beta_{\nu}\Delta}$,
and $\langle J_{S,w}\rangle=\langle J_{H,w}\rangle$, consistent with the observation that
at zero temperature the dissipative dynamics in the SS and the SB models agree \cite{Hanggi,MakriSpin}.
At high temperatures the models support different transport behavior: 
Using the Ohmic form in the wide-band limit,  Eq. (\ref{eq:J2}) with
$\omega_c\gg\Delta$, we get that
%
\bea
\langle J_{H,w}\rangle \xrightarrow{ T_a\gg\Delta} 2\pi\left(\frac{\Delta}{2}\right)^2\frac{\alpha_L\alpha_R}{\alpha_LT_L+\alpha_RT_R}   \delta T,
\nonumber\\
\langle J_{S,w}\rangle \xrightarrow{T_a\gg\Delta}2\pi\left(\frac{\Delta}{2}\right)^3\frac{\alpha_L\alpha_R}{\alpha_L+\alpha_R}  \frac{\delta T}{T_LT_R},
\label{eq:weak}
\eea
with $\delta T=T_L-T_R$.
It is evident that at weak coupling the SB model can
support the thermal diode effect, unlike the SS model, and 
allow for higher currents.
Moreover, Eq. (\ref{eq:weak}) predicts the scaling 
$\langle J_{H,w}\rangle \propto \Delta^2$, in agreement with the (non-adiabatic) factor of NIBA,  see
Eq. (\ref{eq:Current}). In contrast, for the SS model $\langle J_{S,w}\rangle \propto \Delta^3$.
This Born-Markov limit fundamentally deviates 
from NIBA predictions comprising a $\Delta^2\omega_c$ prefactor, see Eq. (\ref{eq:gaussJS}).
This discrepancy may be associated to a recent result showing that in the spin-spin-bath model
the dissipative dynamics predicted by the Redfield equation does not conform with
NIBA \cite{Lu}.


\begin{figure}[htbp]
\vspace{0mm} \hspace{0mm}
{\hbox{\epsfxsize=80mm \epsffile{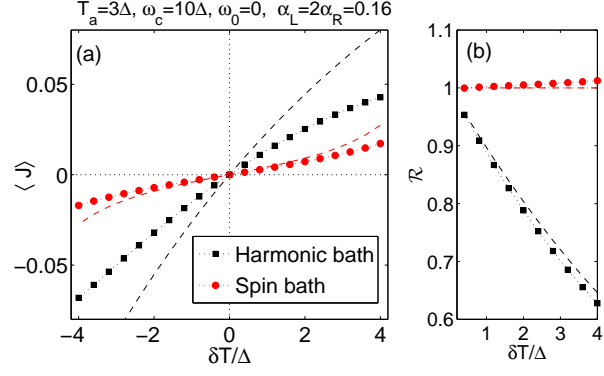}}}
\caption{
Weak coupling limit.
(a) Energy current in the SB  ($\square$) and the SS ($\circ$) models with
the dashed lines calculated from the Born-Markov theory, Eqs. (\ref{eq:weakS}) and
(\ref{eq:weakH}).
(b) Rectification ratio in the SS and SB models using NIBA (symbols) and the Born-Markov theory (dashed lines).
Parameters appear in the figure.
}
\label{FigR1}
\end{figure}

\begin{figure}[htbp]
\vspace{0mm} \hspace{0mm}
{\hbox{\epsfxsize=80mm \epsffile{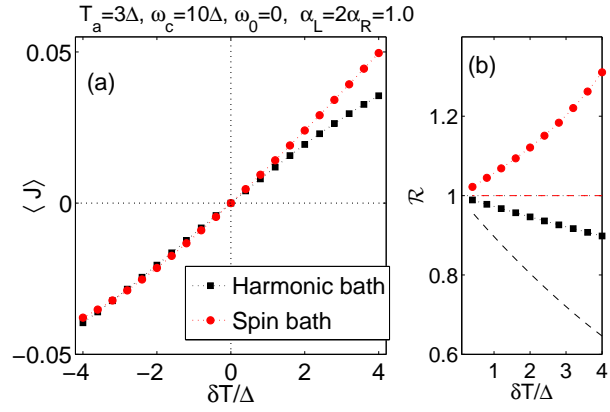}}}
\caption{
Strong coupling limit.
(a) Energy current in the SB ($\square$) and the SS  ($\circ$) models with
the dashed lines calculated from the Born-Markov theory, Eqs. (\ref{eq:weakS}) and
(\ref{eq:weakH}).
(b) Rectification ratio using NIBA (symbols) and the Born-Markov theory (dashed lines).
Parameters appear in the figure.
}
\label{FigR1s}
\end{figure}

\begin{figure}[htbp]
\vspace{0mm} \hspace{0mm}
{\hbox{\epsfxsize=75mm \epsffile{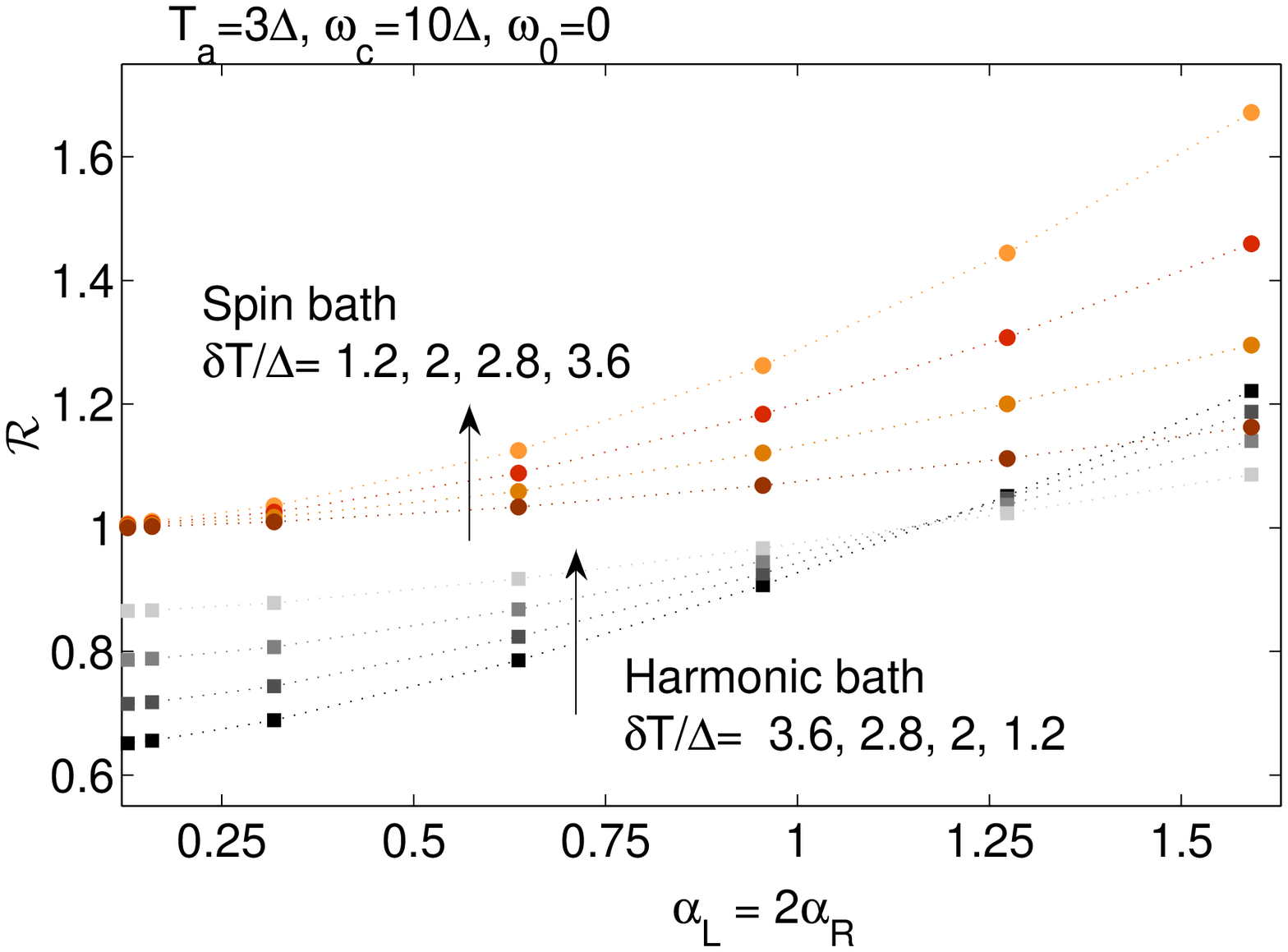}}}
\caption{
Rectification ratio in the SS  ($\circ$) and the SB ($\square$) models. Parameters appear in the figure.
}
\label{FigR2}
\end{figure}

\subsection{Simulations}
\label{simulation}

Simulations were conducted utilizing the same Ohmic spectral density function (\ref{eq:J2}) for the spin bath and the harmonic bath.
We worked in the non-adiabatic limit
with $\Delta/\omega_c\ll1$, and considered unbiased levels $\omega_0=0$.
In simulations where asymmetry was introduced,  we (arbitrarily) set $\alpha_L> \alpha_R$,
but assumed an identical cutoff frequency.
The temperature difference was applied in a symmetric manner with
$T_L=T_a+\delta T/2$,  $T_R=T_a-\delta T/2$.
The current was evaluated numerically from Eq. (\ref{eq:Current}), and it was attained by using Eqs. (\ref{eq:QS}) and (\ref{eq:QH}) 
for the SS and the SB models, respectively. 
The energy current displayed is missing the  $\Delta^2/\hbar$ factor. If one selects $\Delta=10$ meV,
an energy current $\langle J \rangle =0.01$, as reported in Fig. \ref{FigJ2}, translates to
1.5 eV/ns.

Fig. \ref{FigJ2} presents the heat current in the SB and the SS models at high temperatures, $T_a=2\Delta$
and small bias $\delta T/T_a=0.1$.
The x-axis corresponds to the dimensionless Kondo parameter $\alpha=\alpha_{\nu}$,
and we display results from both NIBA and the Born-Markov limit, Eqs. (\ref{eq:weakS}) and (\ref{eq:weakH}).
In the weak-coupling limit the correlation function of the SS model slowly decays,
and NIBA simulations are difficult to converge. We thus present data only for $\alpha>0.15$.

We find that both SS and the SB models predict an enhancement of the current with $\alpha$, at small $\alpha$,
but the SS junction poorly conducts in comparison to the SB model.
At large enough coupling the current decays with $\alpha$, but the turnover occurs early on in the
SB model. The turnover characteristics are predicted by the Marcus expressions described in Sec. \ref{Analytics}. 
Interestingly, at strong coupling the SS junction supports higher currents than the SB system.

In Fig. \ref{FigJ2low} we simulate the heat current at lower temperatures,
$T_a=0.5\Delta$. We find that the SS and the SB systems similarly conduct at strong
coupling, while for $\alpha<0.5$ the SB model supports larger currents.
In this case, the Born-Markov approximation provides only qualitatively-correct results  (SB model).

The current-temperature characteristics are displayed in Fig. \ref{Figd0}, for small $\alpha$,
showing a linear trend close to zero bias ($\delta T=0$) and nonlinearities for $\delta T/\Delta>1$.
The comparison of NIBA to the Born-Markov limit  
reveals a qualitative agreement.
In Fig. \ref{FigR1} we turn to the  asymmetric, $\alpha_L>\alpha_R$, case, 
and we note on marked differences between the models at weak coupling:
While in the SB case a diode-like behavior is established
$\mathcal R\equiv |\langle J(\delta T)\rangle/ \langle J(-\delta T)\rangle|\neq 1$
\cite{SegalRectif,Claire}, the SS model does not support this effect, and the energy current remains
symmetric with $\delta T$.
In the strict weak-coupling limit this phenomenon has been discussed in Ref. \cite{Claire}, and it can be explained
based on Eqs. (\ref{eq:weakS}) and (\ref{eq:weakH}): The system can rectify heat when the statistics of the
baths and the subsystem differ, with spatial asymmetries included.
Furthermore, our analytical expressions of Sec. \ref{Analytics} 
reveal that the unbiased
SS model cannot support significant thermal
rectification in the small-$\alpha$ regime
and at high temperatures, when $\alpha (\omega_c/T_a)^2\ll1$.
It is interesting to note that while the Born-Markov expressions miss the correct values
for the current by up to a factor of 2, see panel (a) in Fig. \ref{FigR1}, the rectification ratio
$\mathcal R$ is well captured within the weak-coupling theory, see  Fig. \ref{FigR1}(b).

Fig.  \ref{FigR1s} displays the current as a function of $\delta T$ for asymmetric setups
$\alpha_L>\alpha_R$ at strong coupling.
We find that the SS junction can in fact rectify heat, but it acts as a better conductor in the direction where
the SB model poorly conducts.
This behavior is explored in more details in  Fig. \ref{FigR2}.
The following observations can be made:
(i) In the SB model the rectification ratio $\mathcal R_{SB}$ does not depend on $\alpha$ at weak coupling,
as expected from Eq. (\ref{eq:weakH}).
However, when $\alpha_L\gtrsim 0.25$,  $\mathcal R_{SB}$ begins to vary with the subsystem-bath coupling parameter. In
parallel, the SS model does not rectify heat at weak coupling showing $\mathcal R_{SS}=1$, and the effect manifests itself
for $\alpha_L\gtrsim0.25$.
(ii) In the SB model the junction better conducts when
the weakly-coupled contact is hot and the strongly-coupled terminal is maintained cold \cite{SegalRectif}.
At a certain large $\alpha$ there is
a special point in parameter space in which the SB junction effectively acts as an harmonic system, with
$\mathcal R_{SB}=1$. Beyond that, the SB junction provides $\mathcal R_{SB}>1$. 
In contrast, the SS junction consistently acts as a better thermal conductor when
the strongly-coupled terminal ($L$) is hotter than the weakly-coupled end ($R$), 
$\mathcal R_{SS}>1$ for $\alpha_L/\alpha_R>1$.

\section{Summary}

In this work we had focused on the out-of-equilibrium spin-spin-bath model, with the central spin coupled to two
separate baths of non-interacting spins. Focusing on the non-adiabatic limit $\Delta<\omega_c$, 
in the first part of the paper we provided the equations of
NIBA with the exact memory kernel.
We showed that spin baths with isolated modes cannot be mapped into the harmonic bath description
through Eq. (\ref{eq:Jeff}), in agreement with early discussions  \cite{Stamp}. In such cases
one should retract to the exact correlation function, 
Eqs. (\ref{eq:R2})-(\ref{eq:R3}).
In the thermodynamic limit, in the so called ``linear-response" regime, the exact discretized kernel
reduces to known results \cite{Makriproof}; in this limit
a spin bath is equivalent to a harmonic bath, only the spectral density function  should be taken to (effectively)
depend on temperature.

In the second part of the paper we considered the linear-response 
thermodynamic limit and compared
the energy transport characteristics of the SB and the SS models in the non-adiabatic regime using the NIBA.
Based on analytic limits and numerical simulations, we pointed on marked, qualitative, differences between the two models:
(i) In the weak-coupling limit harmonic junctions conduct better than the corresponding anharmonic setups.
This result is supported by Born-Markov calculations.
(ii) At weak coupling, $\alpha_{L,R}<0.2$, and in the presence of spatial asymmetries,
the SB junction can rectify heat while the SS model displays symmetric current-temperature bias characteristics.
(iii) At stronger coupling  both the SB and the SS models rectify heat. However, while in the SS model
the current is larger when flowing in the direction of decreasing coupling strength ($T_L>T_R$ and $\alpha_L>\alpha_R$),
the SB model exhibits more complex trends.
(iv) Another striking difference concerns the effect of NDTC:
It is missing  in the unbiased and symmetric SB model, but it robustly shows in the corresponding, unbiased and symmetric, SS model.

Transport characteristics may be employed to explore the properties of the attached terminals,
e.g., the domination of the baths' anharmonic modes in the heat transport process.
From the other way around, specific functionalities
could be engineered and controlled by employing distinct terminals. Particularly, our work suggests that
spin baths (magnetic media, spin glasses) could serve as a good media for supporting 
a strong NDTC through a quantum subsystem, e.g., a qubit. However, such a junction
performs rather poorly as a thermal diode at weak-to-intermediate subsystem-environment coupling.

It would be interesting to extend our work and study thermal conduction problems
with exact techniques, iterative influence functional path integral approaches \cite{MakriSpin,SegalQ},
and complementary perturbative theories,
Green's function methods \cite{Green}. Further, it is important to consider more-involved models for the reservoirs
and mimic complex and realistic environments  \cite{MakriAn,ThossAn,Stamp}: to consider baths
with both harmonic and anharmonic components \cite{StampCPL}, and 
allow for interactions between sub-units in the bath \cite{Kosloff,StampC}.

\begin{acknowledgments}
Support from an NSERC discovery grant is acknowledged.
\end{acknowledgments}

\renewcommand{\theequation}{A\arabic{equation}}
\setcounter{equation}{0}  

\section*{Appendix A: NIBA for the spin-boson model}

The Hamiltonian of the two-bath spin-boson model is given by
\bea
H_{SB}&=&\frac{1}{2}\omega_0 \sigma_z +\frac{1}{2}\Delta \sigma_x +
\sum_{\nu,j}\omega_j b_{\nu,j}^{\dagger}b_{\nu,j}
\nonumber\\
&+&\frac{1}{2}\sigma_z\sum_{\nu,j}
\lambda_{\nu,j}(b^{\dagger}_{\nu,j}+b_{\nu,j}),
\label{eq:HH}
\eea
$\sigma_x$ and $\sigma_z$ are the Pauli matrices, $\omega_0$ is the energy gap between the TLS levels, 
$\Delta$ is the tunneling energy. The reservoirs ($\nu=L,R$) include a collection of non-interacting harmonic
oscillators, $b_{\nu,j}^{\dagger}$ ($b_{\nu,j}$) are the bosonic creation (annihilation) operators of the mode
$j$ in the $\nu$ reservoir. Each mode is coupled to the polarization of the central two-state system with a strength $\lambda_{\nu,j}$.

The SB Hamiltonian (\ref{eq:HH}) can be transformed into the basis of displaced oscillators using the
small polaron transformation \cite{Mahan}, $\tilde H_{SB}=U^{\dagger}H_{SB}U$, where
$U=e^{i\sigma_z\Omega/2}$. The new Hamiltonian reads
%
\bea
\tilde H_{SB} = \frac{\omega_0}{2} \sigma_z +
\frac{\Delta}{2} \left( \sigma_+ e^{i\Omega} + \sigma_- e^{-i\Omega} \right)
+\sum_{\nu,j}\omega_j b_{\nu,j}^{\dagger}b_{\nu,j},
\label{eq:HSBs}
\eea
where $\sigma_{\pm}=\frac{1}{2}(\sigma_x\pm i \sigma_y)$ are the auxiliary Pauli matrices, $\Omega=\sum_{\nu}\Omega_{\nu}$, and $\Omega_{\nu}=i\sum_{j}\frac{\lambda_{\nu,j}}{\omega_{j}}(b_{\nu,j}^{\dagger}-b_{\nu,j})$.
Under the NIBA approximation \cite{Legget,Dekker,Aslangul} generalized to the two-baths case \cite{FR2}, the spin
polarization obeys a time-convolution master equation as in Eq. (\ref{eq:P2}) with the correlation function
\bea
C_H(t)&\equiv&\langle e^{i\Omega(t)}e^{-i\Omega(0)}\rangle
\nonumber\\
&=&
e^{-\sum_{\nu,j}\frac{\lambda_{\nu,j}^2}{\omega_j^2} 
\left[ (1- \cos(\omega_j t)) \coth \left( \frac{\beta_{\nu}\omega_j}{2} \right)  + i \sin(\omega_j t)  \right] }
\nonumber\\
&\equiv& C_{H,L}(t)C_{H,R}(t).
\label{eq:CH}
\eea
Here $C_{H,\nu}\equiv e^{-Q_{H,\nu}(t)}$. Utilizing the definition of the spectral density function
$J_{H,\nu}(\omega)=\pi\sum_{j}\lambda_{j,\nu}^2\delta(\omega-\omega_j)$, we
identify the complex function $Q_{H,\nu}(t)= Q'_{H,\nu}(t)+Q''_{H,\nu}(t)$ as
\bea
Q'_{H,\nu}(t)& = & \int_{0}^{\infty}d\omega\frac{J_{H,\nu}(\omega)}{\pi\omega^2}[1-\cos(\omega t)] [1+2n_B^{\nu}(\omega)],
\nonumber\\
Q''_{H,\nu}(t)& = &  \int_{0}^{\infty}d\omega \frac{J_{H,\nu}(\omega)}{\pi\omega^2}\sin(\omega t),
\label{eq:QH}
\eea
with $n_B^{\nu}(\omega)=[e^{\beta_{\nu}\omega}-1]^{-1}$ as the Bose-Einstein distribution function.

\end{document}